\documentclass{kluwer}
\newdisplay{guess}{Conjecture}
\usepackage{amssymb}
\newif\ifAMStwofonts
\AMStwofontstrue
\def\ee #1 {\times 10^{#1}}
\def\ut #1 #2 { \, \rmn{#1}^{#2}}
\def\u #1 { \, \rmn{#1}}

\def\half{{\textstyle \frac{1}{2}}}
\def\thalf{{\textstyle{ 3\over 2}}}
\let\grad=\nabla
\def\cross{\mathbf{\times}}
\def\curl #1 {\grad \cross #1}
\def\div #1 {\grad \cdot #1}

\def\e{\mathbf{e}}
\def\v{\mathbf{v}}

\def\B{\mathbf{B}}
\def\E{\mathbf{E}}            % E

\def\Bh{\mathbf{\hat{B}}}
\def\fh{\mathbf{\hat{\phi}}}  % phi
\def\rh{\mathbf{\hat{r}}}     % r
\def\vr{v_{r}}
\def\vf{v_{\phi}}            % v_phi
\def\vk{v_{K}}               % v_K

\def\Epa{\mathbf{E'_\parallel}}  % E'_||
\def\Epe{\mathbf{E'_\perp}}  % E'_perp
\def\J{\mathbf{J}}

\def\dE{\mathbf{\delta\E}}
\def\dB{\mathbf{\delta\B}}

\newcommand{\delt} [1] {\frac{\partial #1}{\partial t}}

\input epsf

\begin{document}
\begin{article}
\begin{opening}
\title{Magnetorotational instability in weakly ionised, stratified 
accretion discs}

\author{Raquel \surname{Salmeron}}
\institute{School of Physics, University of Sydney, NSW 2006, Australia}

\author{Mark \surname{Wardle}}
\institute{Physics Department, Macquarie University, NSW 2109, Australia}

\runningauthor{Salmeron \& Wardle}
\runningtitle{Magnetorotational instability in accretion discs}

\date{2003 April 21}
\begin{abstract}
We present a linear analysis of the vertical structure and growth of the 
magnetorotational instability in weakly ionised, stratified accretion discs.
The method includes the effects of the magnetic coupling, the conductivity 
regime of the fluid and the strength of the magnetic field, which is initially 
vertical. The conductivity is treated as a tensor and assumed constant with 
height. 

The Hall effect causes the perturbations to grow faster and act over a much 
more extended section of the disc when the magnetic coupling is low. As a 
result, 
significant accretion can occur closer to the midplane, despite the 
weak magnetic coupling, because of the high column density of the fluid. This 
is an interesting alternative to the commonly held view that accretion 
is relevant mainly in the surface regions of discs, which have a better 
coupling, but a much lower fluid density. 

\end{abstract}
\end{opening}

\section{Introduction}
The magnetorotational instability (MRI) (Balbus \& Hawley 1991(BH91 
hereafter), 
Hawley \& Balbus 1991) transports angular momentum radially 
outwards in accretion discs through the 
distortion of the magnetic field lines that connect fluid elements. 
In protostellar discs, low conductivity is important, specially in the inner 
regions (Gammie 1996; Wardle 1997). As a result, low $k$ modes are relevant
and vertical stratification is a key factor of the analysis. However, most 
models of the MRI in these environments have 
adopted either the ambipolar diffusion or resistive approximations and have 
not simultaneously treated stratification and Hall conductivity. 
We present here a linear analysis of the MRI, including the Hall 
effect, in a stratified disc. 
  
\section{Formulation}
\label{sec:formulation}
  
We write the equations of non-ideal MHD about a local keplerian frame 
corotating with the disc at the angular frequency $\Omega$. The velocity 
field is expressed as a departure from exact keplerian motion, 
$\mathbf{v} = \mathbf{V} - \v_K$, where $\mathbf{V}$ is the velocity 
in the standard laboratory frame $(r, \phi, z)$ and $\v_K$ is the keplerian 
velocity at the radius $r$. We assume
the fluid is weakly ionised, meaning that the abundances of charged species 
are low enough that their inertia and thermal pressure can be neglected, 
together with the effects of ionisation and recombination 
processes in the neutral gas. Under these conditions, separate equations of 
motion for the charged species are not required and the fluid equations are:

\begin{equation}
\delt{\rho} + \div(\rho \v) = 0 \,,	
	\label{eq:continuity}
\end{equation}

\begin{equation}
\delt{\v} + (\v \cdot \grad)\v -2\Omega \vf \rh + \half\Omega \vr\fh 
-\frac{\vk^2}{r}\rh +
\frac{c_s^2}{\rho}\grad\rho +\grad \Phi = \frac{\J\cross\B}{c\rho}\,,
	\label{eq:momentum}
\end{equation}

\begin{equation}
\delt{\B} = \curl (\v \cross \B) - c \curl \E' 
-\thalf \Omega \B_r \fh \,.
	\label{eq:induction}
\end{equation}

\noindent
In the equation of motion (\ref{eq:momentum}), $\Phi$ is the gravitational 
potential for a non self-gravitating disc
and $\v_K^2/r$ is the centripetal term generated by keplerian 
motion. Coriolis terms $2\Omega \vf\rh$ and $\half \Omega 
\vr\fh$ are associated with the use of a local keplerian 
frame and $c_s$ is the isothermal sound speed.
In the induction equation (\ref{eq:induction}), $\E'$ is the electric field 
in the 
frame comoving with the neutrals and $\thalf \Omega \B_r \fh$ accounts for the
generation of toroidal by differential rotation. Additionally, the magnetic 
field must satisfy the constraint
$\div \B = 0$ and the current density must satisfy Ampere 
and Ohm's laws. 
Following Wardle \& Ng (1999), Wardle 1999 (W99) and references 
therein, the current density is expressed as,

\begin{equation}
	\J = \mathbf{\sigma}\cdot \E' = \sigma_{\parallel} \Epa + 
	\sigma_1 \Bh \cross \Epe + \sigma_2 \Epe  \,,
	\label{eq:J-E}
\end{equation}
\noindent

where $\Epa$ and $\Epe$ are the components of $\E'$ parallel and 
perpendicular to $\mathbf{B}$. Note that the conductivity is a 
tensor with 
components $\sigma_{\parallel}$, parallel to the magnetic field, $\sigma_1$, 
the Hall conductivity, and $\sigma_2$, the Pedersen conductivity.

Our model includes vertical stratification, but it is local in 
the radial direction.  The vertical density distribution in hydrostatic 
equilibrium is
$\rho/\rho_o= \exp (-z^2/2H^2)$
where $\rho_o$ is the midplane gas density and 
$H = c_s/\Omega$ is the scaleheight of the disc. 
Equations (\ref{eq:continuity}) to (\ref{eq:J-E}) were linearised about an 
initial steady state where 
$\J = \v = \E' = 0$ and $\B = B \hat{z}$.  Note that as $\E'$ vanishes in the 
initial state, the changes in the conductivity due to the MRI are not 
relevant in this linear formulation and only the unperturbed 
values of $\mathbf{\sigma}$ are required.
Taking perturbations of the form 
$\mathbf{q} = \mathbf{q}_{0} + \mathbf{\delta q}(z) \e^{i\omega t}$ and 
assuming $k = k_z$ we obtained a system of ordinary 
differential equations (ODE) in $\dE$ (the 
perturbations of the electric field in the laboratory frame), $\dB$, and 
the growth rate $\nu=i\omega/\Omega$. Three parameters 
control the evolution of the fluid.
($1$) $v_A/c_s$, the ratio of the Alfv\'en 
speed and the isothermal sound speed of the gas at the midplane, which  
is a measure of the strength of the magnetic field. 
($2$) $\chi_o$, the ratio of the frequency above which 
flux-freezing breaks down and the dynamical frequency of 
the disc. When $\chi_o < 1$ the magnetic field is poorly coupled to the disc. 
($3$) $\sigma_1/\sigma_2$, the 
ratio of the conductivity terms perpendicular to the magnetic field, 
which characterises the \emph{conductivity regime} of the fluid.

We present here results associated with the \emph{ambipolar diffusion} limit 
($\sigma_1=0$), both \emph{Hall} limits ($\sigma_2=0$, $\sigma_1B_z>0$ and 
$\sigma_1B_z<0$) and the \emph{full conductivity} regime 
($\sigma_1=\sigma_2$). 
Ambipolar diffusion is dominant at relatively low densities, when the magnetic 
field is frozen into the ionised component of the fluid. Conversely, Hall 
limits are predominant at intermediate densities, and are characterised by a 
varying degree of magnetic coupling amongst charged species. Typically 
electrons are well coupled, but ions and especially grains tend to be tied to 
the neutrals via collisions. We note that the \emph{Ohmic} regime, which 
dominates at low 
densities, when the charged species are attached to the neutrals via 
collisions, is identical to the ambipolar diffusion regime for the modes of 
interest here.  
We integrated this system of equations vertically as a 
two-point boundary value problem for coupled ODE
with boundary conditions $\delta B_r = \delta B_{\phi} = 0$ and 
$\delta E'_r=1$ at $z=0$ 
and $\delta B_r = \delta B_{\phi} = 0$ at $z/H=5$. 
We refer the reader to Salmeron \& Wardle (in press), for full details of the 
method and results. 

\section{RESULTS}
\label{sec:results}

\begin{figure}
\centerline{\epsfxsize=10.5cm \epsfbox{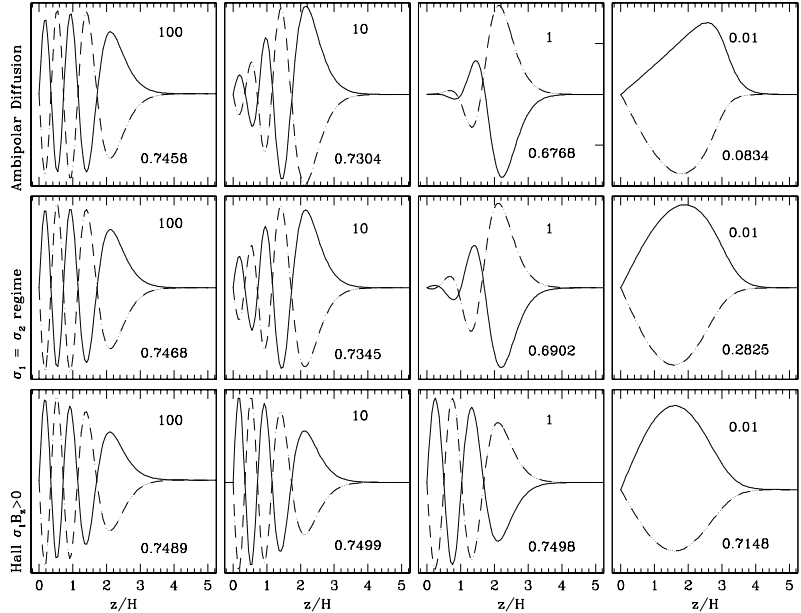}}\vskip 0cm
\caption{Structure and growth rate (shown in the lower right corner of 
each panel) of the most unstable modes of the MRI for 
different conductivity regimes as a function of $\chi_o$ (indicated 
at the top right corner). In all cases $v_A/c_s = 0.1$.}  
\label{fig:1} 
\end{figure}

We compare the structure and growth rate of MRI perturbations as a 
function of the magnetic coupling $\chi_o$ for different conductivity regimes 
and $v_A/c_s=0.1$ (Fig. \ref{fig:1}). At very high magnetic 
coupling ($\chi_o \approx 100$), ideal MHD holds and 
all conductivity regimes are alike (see leftmost column of the figure). 
For a weaker coupling, such that $\chi_o>v_A/c_s$, the structure of the 
ambipolar diffusion limit and full conductivity regimes is similar, signalling 
that ambipolar diffusion is dominant in this region of parameter space. Note 
that both perturbations peak at the node closest to the surface.
This occurs because the maximum growth rate of ambipolar diffusion 
perturbations increases with the local $\chi$ (W99), which in turn is a 
function of height. As a result, at higher $z$ the local growth of the 
instability 
increases, driving the amplitude of these global perturbations to increase. 
Hall perturbations, on the other hand, peak at the node closest to the 
midplane, as $\nu_{max}$ is 
the same  for all $\chi$ (W99) in this case and the instability is not 
driven from any particular vertical location.  
Finally, for $\chi_o<v_A/c_s$ (rightmost column), ambipolar diffusion is 
damped (W99) and the MRI is driven by the Hall effect. In this case ambipolar 
diffusion and full conductivity perturbations differ, the structure of the 
latter ones resembling the Hall limit, as expected.

\begin{figure}
\centerline{\epsfxsize=9.7cm \epsfbox{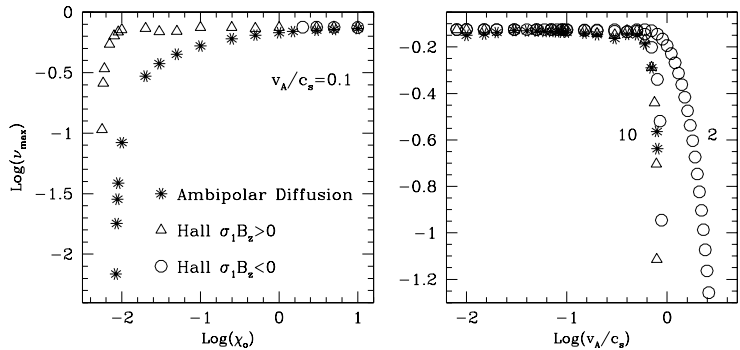}}\vskip 0cm
\caption {Maximum growth of the MRI. Left: as a function of $\chi_o$ with 
$v_A/c_s=0.1$. 
Right: as a function of $v_A/c_s$ for $\chi_o=10$ 
and $\chi_o=2$ (Hall $\sigma_1B_z<0$ regime only)}
\label{fig:2}
\end{figure}

We also explored the growth rate of the perturbations in parameter space. 
Fig. \ref{fig:2} (left panel) shows the growth rate of the most unstable 
perturbations ($\nu_{max}$) as a 
function of $\chi_o$ for ambipolar diffusion and both Hall limits with 
$v_A/c_s=0.1$.
At good coupling the instability grows at $\sim 0.75 \Omega$ in all cases. 
As $\chi_o$ diminishes,  $\nu_{max}$ is reduced at a 
rate that depends on the conductivity regime. The maximum growth rate departs 
significantly from the ideal value for $\chi_o \sim 0.1$ (ambipolar diffusion 
regime) and for $\chi_o \sim 0.01$ (Hall $\sigma_1B_z>0$ limit). The Hall 
$\sigma_1B_z<0$ case could not be analysed for $\chi_o<2$ because in this 
region of parameter space all wavenumbers grow (W99) and our code fails to 
converge.  
These results are consistent with the finding that ambipolar diffusion 
modes grow when $\chi \gtrsim v_A/c_s$ while Hall cases require 
$\chi \gtrsim {v_A}^2/{c_s}^2$ (W99). As a result, for $\chi_o<v_A/c_s$ 
ambipolar diffusion 
perturbations have negligible growth, but when Hall conductivity is 
present the instability still grows at $\nu = 0.2 - 0.3$ (see Fig. 
\ref{fig:1}). In fact, we found that when Hall terms 
dominate, $\nu \sim 0.75$ for $\chi_o \sim 10^{-4}$ even at $v_A/c_s 
\sim 0.01$, as expected.
 
Fig. \ref{fig:2} (right panel) shows the maximum growth rate as 
a function of $v_A/c_s$ for
$\chi_o=10$ and $2$ (the latter one for Hall $\sigma_1B_z<0$ regime only). 
For very weak $v_A/c_s$, the MRI grows at close to the ideal rate in all 
regimes. Generally, increasing the strength of the magnetic field has little 
effect on $\nu_{max}$ until $v_A/c_s \sim 1$, when all perturbations are 
damped 
(e.g. see Fig. \ref{fig:2}, $\chi_o=10$). At this $v_A/c_s$, the wavelength of 
the most unstable modes becomes $\sim H$, the scaleheight of the disc (BH91).
An exception to this occurs in the Hall $\sigma_1 B_z<0$ regime for 
$\chi_o=2$. In this case we found unstable modes for $v_A/c_s$ up to $2.9$. We 
know from the local analysis (W99) that once the local $\chi<2$, unstable 
modes exist for every $kv_A/\Omega$ in this regime. As a result, even for 
suprathermal fields ($v_A/c_s>1$), there are still unstable modes  with 
$kH \lesssim 1$ growing within the disc.
 
\section{Discussion}
\label{sec:discussion}
 
\begin{figure}
\centerline{\epsfxsize=9.5cm \epsfbox{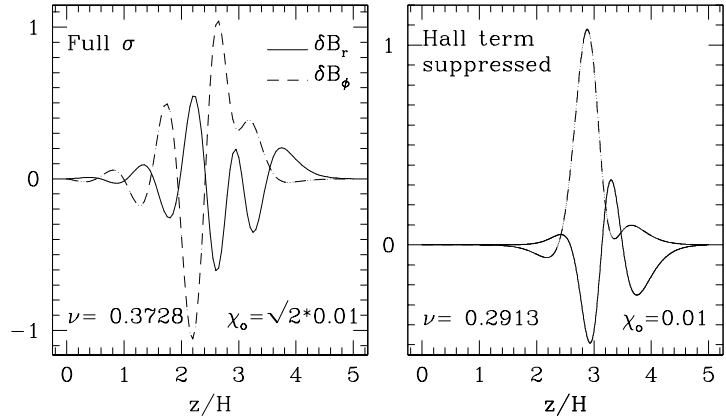}}\vskip 0cm
\caption {Structure and growth rate of the MRI when the 
Hall regime is dominant close to the midplane and ambipolar diffusion 
dominates near the surface (left panel) and under the
ambipolar diffusion approximation (right panel). $v_A/c_s = 0.01$.}
\label{fig:real_discs}
\end{figure}

In a real disc, different conductivity regimes are expected to dominate
at different heights. Despite this, it is common to neglect the 
contribution of Hall terms in studies of 
non-ideal accretion discs, which generally adopt either the ambipolar 
diffusion or 
resistive approximations. To illustrate the way Hall conductivity can 
affect the dynamics and evolution of the fluid we modelled the MRI 
with a set of parameters such that the Hall regime dominates 
closer to the midplane while ambipolar diffusion is predominant near the 
surface, as expected to occur in a real disc. We compared these results 
with the ambipolar diffusion approximation (see Fig. \ref{fig:real_discs}). 
Note that in the full conductivity case
the extent of the dead zone is reduced 
and the growth rate is increased in relation to the ambipolar diffusion 
limit.  This has implications
for the global evolution of the disc, specially considering that even 
though the MRI is damped in the dead zone, velocity fluctuations persist, 
driven by turbulence in the active zones,  and transport angular momentum 
through non-axisymmetric density waves (Stone \& Fleming 2003). 
On the other hand, a quiescent zone would allow dust grains to settle towards 
the midplane and begin to assemble into planetesimals (e.g Weidenschilling \& 
Cuzzi 1993), so the process of planet formation could also 
be affected by Hall conductivity.

Clearly, more detailed modelling is required. The assumption of 
constant conductivity is unrealistic and will affect the 
nature of the global modes. An analysis including a 
$z$-dependent conductivity is underway to examine more fully the 
MRI in low conductivity discs and 
to quantify the importance of Hall terms.

\end{article}
\end{document}